\documentclass[twocolumn]{aastex631}

\usepackage{amsmath}
\usepackage{bm}
\usepackage{xcolor}
\newcommand{\ksp}{\kappa_{\rm Sp}}
\newcommand{\freq}{f_{\rm req}}
\newcommand{\Mdot}{\dot{M}}
\newcommand{\Gcf}{\mathcal{G}_{\rm cf}}
\newcommand{\Xicf}{\Xi_{\rm cf}}
\newcommand{\Ftwentyfive}{\ensuremath{0.055^{+0.009}_{-0.008}}}
\newcommand{\Finner}{\ensuremath{0.18^{+0.09}_{-0.05}}}
\newcommand{\Fouter}{\ensuremath{0.05^{+0.01}_{-0.01}}}

\newcommand{\Tinner}{\ensuremath{64^{+30}_{-20}}}
\newcommand{\Touter}{\ensuremath{75^{+30}_{-19}}}

\newcommand{\Xiinner}{\ensuremath{0.32^{+0.15}_{-0.10}}}
\newcommand{\Xiouter}{\ensuremath{0.37^{+0.15}_{-0.09}}}

\newcommand{\XiouterSystematicRange}{\ensuremath{0.23\text{--}0.52}}

\newcommand{\Qresult}{\ensuremath{0.56^{+0.05}_{-0.04}}}

\newcommand{\Gouter}{\ensuremath{0.261^{+0.034}_{-0.034}}}

\newcommand{\CalInner}{\ensuremath{1.13}}
\newcommand{\CalOuter}{\ensuremath{0.99}}
\newcommand{\CalInnerRange}{\ensuremath{1.02\text{--}1.28}}
\newcommand{\CalOuterRange}{\ensuremath{0.94\text{--}1.04}}
\shorttitle{Isotropic Conduction in Ophiuchus}
\shortauthors{Mitra et al.}

\begin{document}

\title{Can Isotropic Thermal Conduction Heat One of the Hottest Cool Cores?}

\author{Samik Mitra}
\altaffiliation{These authors contributed equally to this work.}
\affiliation{International Centre for Theoretical Sciences, Tata Institute of Fundamental Research, Bengaluru, Karnataka 560089, India}
\email{samik.mitra@icts.res.in}
\correspondingauthor{Samik Mitra}

\author{Ramananda Santra}
\altaffiliation{These authors contributed equally to this work.}
\affiliation{International Centre for Theoretical Sciences, Tata Institute of Fundamental Research, Bengaluru, Karnataka 560089, India}

\author{Norbert Werner}
\affiliation{Department of Theoretical Physics and Astrophysics, Faculty of Science, Masaryk University, Kotl\'{a}\v{r}sk\'{a} 267/2, 611 37 Brno, Czech Republic}

\begin{abstract}
Thermal conduction can act as a heating mechanism for cool-core clusters, yet the same transport must erode any temperature discontinuity it crosses. Ophiuchus permits both effects to be tested in the same atmosphere. Its temperature rises from about $1$ keV in the innermost core to $9$ keV at $r\sim30$ kpc, making conductive heating especially favorable, while two sharply resolved cold fronts independently constrain the same transport. \textcolor{black}{We construct a steady radial energy balance from deprojected \textit{Chandra} density and temperature profiles, with XRISM-based limits on turbulent heating and a subsonic inflow contribution.} Closing the balance with conduction requires only sub-Spitzer coefficients --- \Ftwentyfive\ at 25 kpc and \Finner\ and \Fouter\ at the inner and outer cold fronts --- so conduction can energetically supply the missing heat. Yet the same coefficients, applied normal to the cold-front surfaces, would broaden the fronts to their observed width limits within \Tinner\ and \Touter\ Myr, short compared to the characteristic sloshing timescales that generate and sustain such interfaces. For a representative 200 Myr residence time, the radial conductivity demanded by the energy balance exceeds the cross-front conductivity permitted by front survival by roughly a factor of three, with the required anisotropy exceeding unity in 99.9\% and 98.8\% of accepted models. The result leaves two broad regimes. Thermal conduction is either subdominant in Ophiuchus or contributes substantially but is strongly anisotropic, with heat transport suppressed across the cold fronts relative to the radial direction. Together they disfavor locally isotropic conduction as a dominant heating mechanism in cool cores generally.
\end{abstract}


\section{Introduction}\label{sec:introduction}

The dense cores of relaxed galaxy clusters radiate on timescales far shorter than the Hubble time, yet the observed rates of cooling and condensation remain well below those expected from unopposed losses \citep{Fabian1994,Peterson2003}. Mechanical feedback from the central active galactic nucleus (AGN) is the leading source of compensating power, through cavities, shocks, uplift, sound waves, and mixing \citep{McNamara2007,Fabian2012,Birzan2004,Rafferty2006}. Electron thermal conduction provides a complementary channel because a positive radial temperature gradient can carry heat inward from the hotter atmosphere. Classical estimates showed that a substantial fraction of the Spitzer conductivity can stabilize a cooling core under favorable conditions \citep{Binney1981,Bertschinger1986,Zakamska2003,Voigt2004}. Whether that heat can actually reach the core, however, depends on magnetic geometry and plasma microphysics. Electrons stream preferentially along magnetic field lines \citep{Braginskii1965}, tangled fields reduce the isotropic average \citep{Narayan2001}, buoyancy instabilities can reorganize field topology \citep{Parrish2005,Quataert2008,Parrish2009,Perrone2022}, and kinetic effects can suppress the parallel flux below its collisional value \citep{RobergClark2016}.

In particular, three-dimensional cool-core calculations show that the heat-flux-driven buoyancy instability (HBI) can rotate magnetic fields away from the radial temperature gradient and reduce the effective radial conductivity to $\lesssim 0.1\,\kappa_{\rm Sp}$, allowing radiative cooling to outrun conductive replenishment in low-entropy cores \citep{Parrish2009}. On larger scales, thermal conduction also leaves a distinct statistical signature by damping density and temperature fluctuations while leaving the turbulent velocity cascade comparatively unchanged, providing a complementary bulk-ICM transport diagnostic \citep{Gaspari2014}.

Cold fronts probe the same transport from a different direction. These contact-like interfaces separate denser, cooler gas from hotter, more rarefied plasma while maintaining approximate pressure continuity \citep{Markevitch2000,Ascasibar2006,Markevitch2007}. In sloshing cores, shear wraps the interfaces around the cluster centre and can stretch magnetic fields tangentially along their surfaces. Heat conduction normal to a front smooths its temperature and density jumps, so the observed width places a direct upper limit on cross-front transport. Such arguments have already implied suppression by one to two orders of magnitude below the Spitzer value in several systems \citep{Ettori2000,Vikhlinin2001,Markevitch2001}. The same conclusion is not confined to contact discontinuities: temperature structure in the bulk of the merging cluster A754 requires an effective conductivity at least an order of magnitude below Spitzer on $\sim100$ kpc scales \citep{Markevitch2003}. Cold-front widths themselves contain additional dynamical information. Idealized moving-front calculations show that advection near the stagnation region can establish a quasi-steady conductive layer whose thickness is set jointly by thermal diffusivity and the flow, rather than by unrestricted diffusion alone \citep{Xiang2007}. Deep observations of the Virgo cold front further show that preserving very sharp density and temperature jumps can require suppression not only across the interface but also along magnetic field lines when compared with anisotropic-conduction MHD calculations \citep{WernerVirgo2016}. MHD calculations provide a natural interpretation in which shear-amplified magnetic layers insulate the cold gas while preserving some radial connectivity through the core \citep{Lyutikov2006,ZuHone2011,ZuHone2013,ZuHone2021}. What those local constraints do not determine is whether the allowed conductivity is large enough to matter for the global cooling budget.

Ophiuchus brings these two questions together in an unusually favorable system. Its cool core lies in one of the hottest and X-ray-brightest nearby clusters, with the temperature rising from $1$--$2$ keV near the centre to several keV across the inner tens of kiloparsecs \citep{Fujita2008,Million2010,Werner2016}. \textit{Chandra} resolves two sloshing cold fronts inside the cooling region. The inner front spans $r\simeq4.5$--10 kpc with density jump $J_n=2.0\pm0.3$ and Gaussian width $w<1.5$ kpc, while the outer front lies near $r\simeq43$ kpc with $J_n=1.7\pm0.1$ and $w<5.3$ kpc \citep{Werner2016}. The same Chandra study already emphasized that Ophiuchus presents a non-trivial conduction problem: its steep core temperature gradient argues against unrestricted heat transport, while the low-amplitude density fluctuations in the hotter outer atmosphere were discussed as potentially being damped by conduction \citep{Werner2016}. Our aim is therefore not to establish that conduction is suppressed, but to determine whether the suppressed transport that remains is energetically sufficient and simultaneously compatible with both cold fronts. The central radio mini-halo further indicates a structured non-thermal environment \citep{Murgia2010,Giacintucci2020,Giacintucci2025,Botteon2025}. XRISM now adds direct spectroscopic constraints on the gas motions, measuring line-of-sight dispersions of $115\pm7$ and $186\pm9\ {\rm km\,s^{-1}}$ together with projected temperatures of $5.8\pm0.2$ and $8.4\pm0.2$ keV in the inner and outer source regions \citep{Fujita2025}. These velocity measurements bound the turbulent contribution to the thermal budget, while a deeper \textit{Resolve} analysis independently finds that turbulent dissipation remains insufficient to offset cooling \citep{Russell2026}.

This combination makes Ophiuchus a direct test of whether energetically relevant conduction can remain locally isotropic. We first determine the radial conductivity required to carry the power left after the modeled turbulent and inflow terms. We then ask whether that same transport strength can act normal to both observed fronts without broadening them beyond their measured limits. The central quantity is therefore not a suppression factor inferred from one interface alone, but the ratio between the conductivity demanded by the core energy budget and the conductivity permitted by front survival in the same atmosphere. Section~\ref{sec:model} defines the thermodynamic, energetic, and front-broadening framework. Section~\ref{sec:results} presents the resulting transport constraint, and Section~\ref{sec:discussion} discusses its physical interpretation and broader implications for cool-core heating.

\section{Physical Framework}\label{sec:model}

\subsection{Thermodynamic atmosphere and projected observables}\label{sec:atmosphere}

The conductive heat flux is controlled by the local temperature gradient, so the transport constraint is only as reliable as the thermodynamic atmosphere from which that gradient is inferred. We construct a spherical-equivalent reference atmosphere from the deprojected sector measurements described by \citet{Werner2016}, retaining their radial bins, sector identities, and asymmetric uncertainties. Let $\Delta\phi_i$ denote the opening angle of sector $i$, with $n_{{\rm e},i}(r)$ and $T_i(r)$ its deprojected electron density and temperature. We define

\begin{equation}
\begin{aligned}
 n_{\rm e,eq}(r)&=\left[\frac{\sum_i\Delta\phi_i n_{{\rm e},i}^2(r)}{\sum_i\Delta\phi_i}\right]^{1/2},\\
 T_{\rm eq}(r)&=\frac{\sum_i\Delta\phi_i n_{{\rm e},i}^2(r)T_i(r)}{\sum_i\Delta\phi_i n_{{\rm e},i}^2(r)},
\end{aligned}
\label{eq:equiv}
\end{equation}

The $n_e^2$ weighting follows the approximate emission-measure dependence of the thermal X-ray emissivity. Equation~(\ref{eq:equiv}) is therefore a spherical representation of the measured sectors rather than an assumption that the core is azimuthally uniform. Because conduction depends nonlinearly on $T$, averaging the atmosphere and evaluating the conductive capacity need not commute. We quantify this effect with $\mathcal R_q\equiv C_{q,\rm eq}/\langle C_{q,i}\rangle_\phi$, where $C_q=\ksp(T)|dT/dr|$ and the brackets denote opening-angle weighting. We make the analogous comparison for $n_e^2\Lambda(T,Z)$ to assess the impact on radiative cooling.

The density profile requires two smooth changes in logarithmic slope to represent both the inner core and the outer line-of-sight atmosphere,
\begin{equation}
\begin{aligned}
 n_e(r)=&\,n_{25}\left(\frac{r}{25\,\mathrm{kpc}}\right)^{-p_0}
 B_1(r)^{-(p_1-p_0)/4}\\
 &\times B_2(r)^{-(p_2-p_1)/4},\quad
 B_j(r)=\frac{1+(r/r_j)^4}{1+(25\,\mathrm{kpc}/r_j)^4},
\end{aligned}
\label{eq:density}
\end{equation}
where $n_{25}$ is the electron-density normalization at 25 kpc, $p_0$, $p_1$, and $p_2$ are the successive logarithmic slopes, and $r_1$ and $r_2$ are the transition radii. The corresponding local positive density slope is
\begin{equation}
\begin{aligned}
 p_{\rm eff}(r)&\equiv-\frac{d\ln n_e}{d\ln r}
 =p_0+(p_1-p_0)\frac{x_1}{1+x_1}
 +(p_2-p_1)\frac{x_2}{1+x_2},\\
 x_j&=(r/r_j)^4.
\end{aligned}
\label{eq:peff}
\end{equation}

The temperature profile must reproduce the cool central gas, the rapid rise through the inner core, and the weaker outer trend needed for the line-of-sight projection. We use
\begin{equation}
\begin{aligned}
 T(r)&=T_0\frac{x+\tau}{x+1}
 \left[1+\left(\frac{r}{r_o}\right)^2\right]^{-c/2},\\
 x&=\left(\frac{r}{r_c}\right)^a,\qquad
 \tau=\frac{T_{\min}}{T_0},
\end{aligned}
\label{eq:temperature}
\end{equation}
where $T_{\min}$ and $T_0$ set the central and hot-core temperature scales, $r_c$ and $a$ control the cool-core rise, and $r_o$ and $c$ allow a weak outer flattening or decline. Since the conductive capacity depends directly on the gradient, we use the analytic logarithmic slope
\begin{equation}
 g(r)\equiv\frac{d\ln T}{d\ln r}
 =a\frac{x(1-\tau)}{(x+\tau)(x+1)}-c\frac{(r/r_o)^2}{1+(r/r_o)^2}.
\label{eq:geff}
\end{equation}

\textcolor{black}{Fitting Equations~(\ref{eq:density}) and (\ref{eq:temperature}) gives the nominal parameters $n_{25}=0.02043\,\mathrm{cm^{-3}}$, $(p_0,p_1,p_2)=(0.8615,0.6755,2.0253)$, $(r_1,r_2)=(14.38,221.93)$~kpc, $T_0=10.53$~keV, $T_{\min}=0.72$~keV, $r_c=18.87$~kpc, $a=1.042$, $r_o=54.81$~kpc, and $c=0.0964$. The logarithmic rms residual of the temperature fit is 0.0241.} A simpler three-parameter saturating form performs comparably over 2--60 kpc, but does not independently control the outer slope. We retain Equation~(\ref{eq:temperature}) because the XRISM projection receives line-of-sight emission from beyond the 50 kpc energy-analysis interval. The final constraints are not tied to the nominal fit. They are evaluated over a profile library generated by perturbing the sector measurements within their uncertainties and by leave-one-sector-out refits, so the allowed thermodynamic gradients propagate into the transport calculation.

We additionally reject thermodynamically pathological profiles with a hydrostatic-like admissibility test. The proxy enclosed mass is
\begin{equation}
 M_{\rm HE}(r)=\frac{k_BT(r)r}{G\mu m_p}\left[p_{\rm eff}(r)-g(r)\right]
\label{eq:hse}
\end{equation}
where $k_B$ is Boltzmann's constant, $G$ is the gravitational constant, $m_p$ is the proton mass, and $\mu=0.61$ is the mean molecular weight per particle. We require $M_{\rm HE}$ to remain positive and non-decreasing between 2 and 50 kpc. This criterion is not used to infer the cluster mass. It only removes combinations of $p_{\rm eff}$ and $g$ that cannot represent a physically admissible atmosphere.

The thermodynamic model is three-dimensional, whereas the XRISM measurements are line-of-sight projections through finite detector regions. We therefore integrate the model through the live-pixel masks of the two \textit{Resolve} source regions rather than comparing each measurement with a single spherical shell. For a detector region $\mathcal R$,
\begin{equation}
 \begin{aligned}
 T_{X,\mathcal R}&=\frac{\int_{\mathcal R} n_e^2\epsilon_X(T,Z)T\,dV}
 {\int_{\mathcal R}n_e^2\epsilon_X(T,Z)\,dV},\\
 \sigma_{\mathcal R}^2&=\frac{\int_{\mathcal R}n_e^2\epsilon_{\rm FeK}(T,Z)\sigma_v^2\,dV}
 {\int_{\mathcal R}n_e^2\epsilon_{\rm FeK}(T,Z)\,dV},
 \end{aligned}
\label{eq:spectral}
\end{equation}
where $\epsilon_X$ and $\epsilon_{\rm FeK}$ are relative emissivity weights for the broad 2--12 keV emission and the Fe-K line complex. We approximate these weights with a varied collisional ionization equilibrium (CIE) surrogate containing bremsstrahlung and Fe~XXV/XXVI kernels \citep{Foster2012}. This is a forward-weighting calculation rather than a response-folded XRISM spectral fit, so we include a conservative 10\% projection/emissivity systematic and require accepted atmospheres to reproduce the two published projected temperatures within the combined tolerance.

The velocity field entering the turbulent-dissipation term is parameterized phenomenologically as
\begin{equation}
 \sigma_v(r)=\sigma_{25}\left(\frac{r}{25\,\mathrm{kpc}}\right)^u,
\label{eq:sigmamodel}
\end{equation}
with $0\le u\le1$. Its normalization is set by the turbulent-energy closure below, not by fitting the two XRISM line widths directly. The closure is nevertheless XRISM anchored. \citet{Fujita2025} inferred a turbulent-heating fraction of order 0.4 from the measured velocity width and an assumed outer scale, while the deeper analysis of \citet{Russell2026} finds turbulent dissipation roughly a factor of three below cooling. We therefore adopt $L_{\rm turb}/L_{\rm cool}=0.36\pm0.09$ at 25 kpc as a conservative upper-limit normalization. If only a fraction $f_\sigma=0.75$--1 of the measured width is associated with genuine turbulence, this limit is reduced as $f_\sigma^3$. The projected $\sigma_v$ profile is then required to remain consistent with the published XRISM widths. Table~\ref{tab:priors} summarizes the principal model inputs.

\begin{table}[!t]
\centering
\caption{Principal inputs and model treatments.}
\label{tab:priors}
\tabletypesize{\scriptsize}
\setlength{\tabcolsep}{1.5pt}
\begin{tabular}{@{}ll@{}}
\hline
Quantity & Treatment \\
\hline
$n_e(r),T(r)$ & \textcolor{black}{Werner et al. (2016) + sector omissions} \\
Energy domain & 2--50 kpc \\
$Z_{\rm in},Z_{\rm out}$ & $0.75\pm0.05$, $0.44\pm0.04$ \\
$L_{\rm turb}/L_{\rm cool}$ & $0.36\pm0.09$ at 25 kpc (XRISM limit) \\
$u,m$ & uniform $0$--1 \\
$l_{25}$ & log-uniform 3--60 kpc \\
$f_\sigma$ & uniform 0.75--1 \\
$\Mdot$ & \textcolor{black}{sampled $0$--$(0.97\pm0.12)\,M_\odot\,\mathrm{yr}^{-1}$} \\
$r_{\rm cf}^{\rm in,out}$ & 4.5--10, $43\pm3$ kpc \\
$J_n^{\rm in,out}$ & $2.0\pm0.3$, $1.7\pm0.1$ \\
$w_{99}^{\rm in,out}$ & 1.5, 5.3 kpc \\
\hline
\end{tabular}
\end{table}

\subsection{Subsonic energy balance, turbulence, and conduction}\label{sec:energy}

The atmosphere fixes both the radiative loss rate and the radial temperature gradient available to transport heat inward. It does not determine how much of the cooling is already offset by other processes. We therefore write the cumulative energy bookkeeping as
\begin{equation}
 L_{\rm cool}=L_{\rm turb}+L_{\rm comp}+L_{\rm cond}+L_{\rm other},
\label{eq:bookkeeping}
\end{equation}
where $L_{\rm cool}$ is the radiative loss, $L_{\rm turb}$ the turbulent dissipation, $L_{\rm comp}$ the net thermal contribution of the slow inflow, $L_{\rm cond}$ the conductive term, and $L_{\rm other}$ all channels not modeled explicitly, including AGN work, mixing, waves, shocks, and cosmic-ray heating. This decomposition is deliberately non-exclusive. The calculation isolates the power left after the modeled turbulent and inflow terms and asks how much conductivity would be required if conduction carried that residual.

For a steady spherical mass flux, with positive $\Mdot$ denoting inflow,
\begin{equation}
\begin{aligned}
 \Mdot&=-4\pi r^2\rho v_r,\\
 \rho v_rT\frac{ds}{dr}&=-n_en_i\Lambda
 -\frac{1}{r^2}\frac{d}{dr}(r^2q_r)+H_{\rm turb},
\end{aligned}
\label{eq:entropy}
\end{equation}
where $v_r<0$ is inferred from the sampled $\Mdot$ and $\rho=\mu_e m_pn_e$ with $\mu_e=1.17$. The mean inflow velocity is distinct from the turbulent dispersion $\sigma_v$ and from the XRISM line-of-sight velocity measurements. The left-hand side advects entropy. The terms on the right describe radiative losses, the divergence of the conductive flux, and turbulent dissipation, with $n_i\simeq0.92n_e$ and $\Lambda(T,Z)$ the CIE cooling function. We verify that the inferred flow remains deeply subsonic through $\mathcal M_r=|v_r|/c_s\ll1$, where $c_s=(\gamma k_BT/\mu m_p)^{1/2}$ and $\gamma=5/3$. The cumulative cooling luminosity is
\begin{equation}
\begin{aligned}
 L_{\rm cool}(<r)&=4\pi\int_{r_{\min}}^r n_e(r')n_i(r')
 \Lambda[T(r'),Z(r')]\,r'^2dr',\\
 r_{\min}&=2\ {\rm kpc},
\end{aligned}
\label{eq:lcool}
\end{equation}
which makes explicit the approximate $n_e^2$ dependence of the emissivity and the resulting sensitivity of the energy budget to the density normalization.

For turbulent dissipation we adopt an order-unity cascade estimate. If isotropic turbulence has $v_{\rm 3D}=\sqrt{3}\,\sigma_v$, then $\rho v_{\rm 3D}^3/l_t\simeq3^{3/2}\rho\sigma_v^3/l_t$. We write
\begin{equation}
 H_{\rm turb}=5\frac{\rho\sigma_v^3}{l_t},\qquad
 l_t(r)=l_{25}\left(\frac{r}{25\,\mathrm{kpc}}\right)^m,
\label{eq:turb}
\end{equation}
where $l_t$ is the turbulent outer scale and the coefficient 5 approximates $3^{3/2}$ multiplied by an order-unity cascade coefficient. The cumulative turbulent power is $L_{\rm turb}(<r)=4\pi\int H_{\rm turb}r'^2dr'$. Let $\eta_{\rm turb,25}$ denote the adopted cumulative turbulent-heating fraction at 25 kpc. Imposing $L_{\rm turb}(<25\,\mathrm{kpc})=\eta_{\rm turb,25}L_{\rm cool}(<25\,\mathrm{kpc})$ gives
\begin{equation}
 \sigma_{25}=\left[
 \frac{\eta_{\rm turb,25}L_{\rm cool}(<25)l_{25}}
 {20\pi\displaystyle\int_{r_{\min}}^{25\,\mathrm{kpc}}
 \rho(r)(r/25\,\mathrm{kpc})^{3u-m}r^2dr}
 \right]^{1/3}.
\label{eq:sigma25}
\end{equation}
Thus $\sigma_{25}$ follows from the atmosphere, the XRISM-anchored turbulent fraction, the radial exponent $u$, the outer scale $l_{25}$, and its radial exponent $m$. It is not fitted to the two \textit{Resolve} widths, but it is not statistically independent of them because the adopted turbulent-heating normalization originates from XRISM. We forward-project the resulting $\sigma_v(r)$ through Equation~(\ref{eq:spectral}) and require the projected values not to exceed the published widths. \textcolor{black}{This is also consistent with theoretical arguments that turbulent dissipation alone would require near-sonic motions to balance radiative cooling in cluster cores, substantially larger than the subsonic motions observed in cool-core systems \citep{Banerjee2014}.}

The slow inflow adds a much smaller advective and compressional contribution. For an ideal gas,
\begin{equation}
 \frac{dL_{\rm comp}}{d\ln r}
 =\frac{\Mdot k_BT}{\mu m_p}
 \left[p_{\rm eff}+\frac{g}{\gamma-1}\right],
\label{eq:compression}
\end{equation}
where the $p_{\rm eff}$ term is direct compressional $P\,dV$ work and the $g/(\gamma-1)$ term describes advection of the internal-energy gradient. Equation~(\ref{eq:compression}) therefore represents the thermal effect associated with the prescribed subsonic inflow. \textcolor{black}{The subsonic-inflow contribution is not an additional term in Equation~(\ref{eq:entropy}). It is the thermal effect associated with the mean entropy-advection term $\rho v_rT\,ds/dr$ on the left-hand side, recast in Equation~(\ref{eq:compression}) as compressional work plus advection of the internal-energy gradient. The mass inflow rate is not measured by XRISM and is not inferred from the velocity widths; instead, we treat $\Mdot$ as a bounded nuisance parameter over the range listed in Table~\ref{tab:priors}. The resulting flow remains deeply subsonic, and its contribution to the thermal budget is correspondingly small.}It is not a solution of the radial momentum equation.

We describe radial electron conduction with a flux-limited Spitzer law,
\begin{equation}
\begin{aligned}
 q_r&=-\frac{f_r\ksp\,dT/dr}
 {1+f_r\ksp|dT/dr|/q_{\rm sat}},\\
 \ksp&=\frac{1.84\times10^{-5}}{\ln\Lambda_C}T^{5/2},
\end{aligned}
\label{eq:flux}
\end{equation}
with Coulomb logarithm $\ln\Lambda_C=37$ and saturated flux
\begin{equation}
 q_{\rm sat}=0.4n_ek_BT\left(\frac{2k_BT}{\pi m_e}\right)^{1/2}.
\label{eq:qsat}
\end{equation}
Here $f_r$ is the effective radial conductivity in units of the classical Spitzer value. The limiter prevents the diffusive expression from exceeding the electron free-streaming scale when the mean free path becomes comparable to the temperature-gradient length. Equation~(\ref{eq:flux}) approaches the ordinary Spitzer form when $f_r\ksp|dT/dr|\ll q_{\rm sat}$ and the saturated limit $|q_r|\sim q_{\rm sat}$ for sufficiently steep gradients \citep{Spitzer1962,Cowie1977,Mitra2023}. The accepted radial solutions remain safely unsaturated, but retaining the limiter keeps the same transport prescription applicable to the sharper cold-front gradients.

After subtracting the explicit turbulent and inflow terms, the remaining cumulative power is
\begin{equation}
\begin{aligned}
 L_{\rm res}(<r)&=L_{\rm cool}(<r)-L_{\rm turb}(<r)-L_{\rm comp}(<r),\\
 \eta_{\rm cond}(r)&=\frac{L_{\rm cond}(<r)}{L_{\rm res}(<r)}.
\end{aligned}
\label{eq:eta}
\end{equation}
The residual $L_{\rm res}$ can therefore be supplied by conduction, by $L_{\rm other}$, or by both. The limiting case $\eta_{\rm cond}=1$ assigns the entire residual to conduction and defines the largest energetically relevant conductive share. Let $F_{\rm res}=L_{\rm res}/(4\pi r^2)$ be the inward heat-flux magnitude required in that limit. Since $dT/dr=gT/r$, inverting Equation~(\ref{eq:flux}) gives the reference spherical Spitzer fraction
\begin{equation}
 \freq(r)=\frac{F_{\rm res}}
 {\ksp(gT/r)\,[1-F_{\rm res}/q_{\rm sat}]}.
\label{eq:fheat}
\end{equation}
The quantity $\freq$ is therefore a requirement, not a measurement of the local conductivity. It is the coefficient needed only if conduction carries the full residual power.

A front-containing sector need not carry the spherical-mean radial heat flux. We write $F_{r,\rm cf}=\Gcf\eta_{\rm cond}F_{\rm res}$ and define the directly relevant product $\Xicf\equiv\Gcf\eta_{\rm cond}$, where $\Gcf$ is the front-sector radial flux relative to the spherical mean. The representative-sector hypothesis is $\Gcf=1$. The corresponding local radial coefficient is
\begin{equation}
 f_r(\Xicf)=\frac{\Xicf F_{\rm res}}
 {\ksp(gT/r)\,[1-\Xicf F_{\rm res}/q_{\rm sat}]}.
\label{eq:frlocal}
\end{equation}
Because $F_{\rm res}/q_{\rm sat}\ll1$ in the accepted atmospheres, this reduces to $f_r\simeq\Xicf\freq$. Equality $f_r=\freq$ is obtained only for $\Xicf=1$, which combines the maximal conductive closure with a representative front sector.

\subsection{Cold-front broadening}\label{sec:frontmodel}

The observed sharpness of each front constrains the conductivity normal to its surface. We reconstruct the temperature contrast from approximate pressure balance rather than treating it as an independently measured temperature jump. If $J_n=n_{e,c}/n_{e,h}$ is the measured density jump, then $T_c=T_h/J_n$, where the subscripts $c$ and $h$ denote the cold and hot sides. Since $\ksp\propto T^{5/2}$, the conductivity averaged across this reconstructed jump is
\begin{equation}
 \bar\kappa_{\rm Sp}=\frac{2\kappa_{\rm Sp}(T_h)}{7}
 \frac{1-(T_c/T_h)^{7/2}}{1-T_c/T_h}.
\label{eq:kappabar}
\end{equation}
An analytic estimate for the time required to broaden an initially sharp interface to Gaussian width $w$ is
\begin{equation}
 t_{\rm an}(f_\perp)=\frac{C_{P,c}w^2}{2f_\perp\bar\kappa_{\rm Sp}}
 \left(1+\frac{f_\perp\bar\kappa_{\rm Sp}\Delta T}{wq_{\rm sat,c}}\right),
\label{eq:broadening}
\end{equation}
where $f_\perp$ is the Spitzer fraction normal to the front, $\Delta T=T_h-T_c$, $q_{\rm sat,c}$ is the cold-side saturated flux, and $C_{P,c}=(5/2)(1.92n_{e,c})k_B$ is the cold-side constant-pressure heat capacity per unit volume. In the unsaturated limit, $t_{\rm an}\propto w^2/f_\perp$, so a narrower observed interface permits less cross-front transport at fixed exposure time.

We calibrate this estimate with a nonlinear one-dimensional diffusion calculation that evolves a pressure-balanced, two-sided interface using the full $T^{5/2}$ conductivity and the same flux limiter. The evolved density profile is fitted with the Gaussian-smoothed step used to define the observed width. The median numerical-to-analytic time corrections are \CalInner\ and \CalOuter\ for the inner and outer fronts, with representative ranges \CalInnerRange\ and \CalOuterRange. The calculation is intrinsic and one-dimensional, so it does not include front curvature, line-of-sight projection, or the \textit{Chandra} point-spread function (PSF). For an assumed residence time $t_{\rm res}$, inversion of the calibrated broadening relation yields the largest permitted normal conductivity $f_\perp^{\max}$, and hence
\begin{equation}
 \Xicf^{\max}(r_{\rm cf})=\frac{f_\perp^{\max}}{\freq(r_{\rm cf})},\qquad
 \eta_{\rm iso,max}=\min\left[1,\frac{\Xicf^{\max}}{\Gcf}\right].
\label{eq:ximax}
\end{equation}
The data directly constrain $\Xicf^{\max}$. Converting that limit into an isotropic conductive share additionally requires a choice of $\Gcf$, which encodes the azimuthal heat-flux geometry.

\begin{figure}[!t]
\centering
\includegraphics[width=\columnwidth]{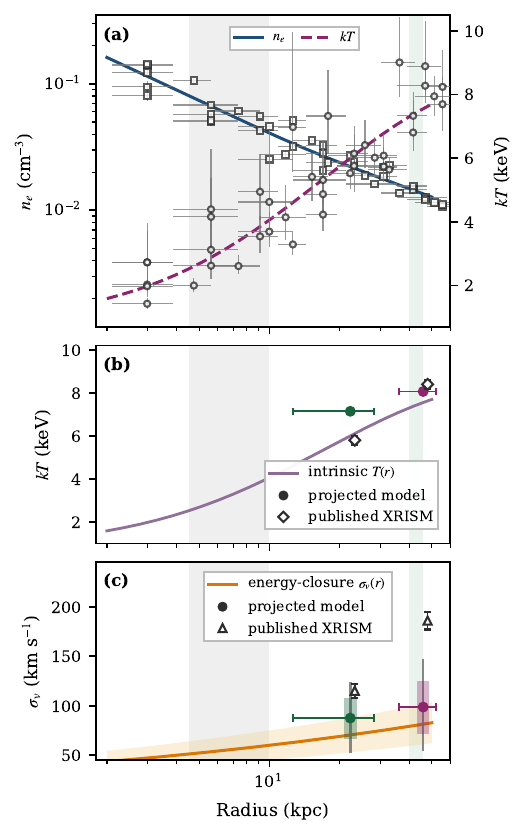}
\caption{Thermodynamic structure and its mapping to XRISM is shown. \textit{Upper panel:} The accepted electron-density and temperature profiles with 68\% envelopes, overlaid on the deprojected five-sector measurements of \citet{Werner2016} is shown. Grey and green bands mark the inner and outer cold fronts in all panels. \textit{Middle panel:} The intrinsic temperature profile together with the line-of-sight projected model values in the two Resolve source masks (horizontal bars), compared with the published XRISM temperatures (open diamonds) is shown. \textit{Lower panel:} The intrinsic $\sigma_v(r)$ profile from the turbulence-energy limits (orange curve), its Fe-K-weighted projected values (points), and the published line widths, which are treated as upper limits on the turbulent component (open triangles) is shown.}
\label{fig:profiles}
\end{figure}

\section{Results}\label{sec:results}

The analysis proceeds in three steps. We first establish that the accepted atmospheres preserve an inward conductive gradient and remain consistent with the XRISM projections. We then determine how much radial conductivity would be required to carry the residual cooling demand. Finally, we compare that requirement with the cross-front conductivity allowed by the two measured interfaces.

Across both cold-front radii the accepted density profiles decrease outward while the temperature rises, so the radial gradient can carry heat toward the centre. The logarithmic temperature slopes are $g_{\rm in}=\Qresult$ and $g_{\rm out}=\Gouter$ (Figure~\ref{fig:profiles}a). The sector-averaging diagnostic gives $\mathcal R_q=0.78$--1.25 between 5 and 50 kpc, with values of 0.78, 0.86, and 1.24 at the inner front, 25 kpc, and the outer front. By comparison, the corresponding cooling-emissivity average agrees with the spherical value to better than $7\times10^{-4}$. The spherical representation therefore changes the conductive capacity at the $\lesssim25\%$ level while leaving the radiative-loss estimate essentially unchanged. This scale is smaller than the full model-sensitivity ranges quoted below and does not create an order-unity transport mismatch by itself.

The same accepted atmospheres are forward-projected through Equation~(\ref{eq:spectral}). Their projected temperatures reproduce the published \textit{Resolve} values (Figure~\ref{fig:profiles}b). The accepted velocity profiles have median $\sigma_{25}\simeq72\ {\rm km\,s^{-1}}$ and $u\simeq0.20$, and their Fe-K-weighted projections remain below the published line widths (Figure~\ref{fig:profiles}c). Because the turbulent normalization is XRISM anchored, this comparison is an internal consistency check on the radial closure and projection rather than an independent prediction of the measured widths.

\subsection{Energy budget and required radial conductivity}\label{sec:res_energy}

\begin{figure}[!t]
\centering
\includegraphics[width=\columnwidth]{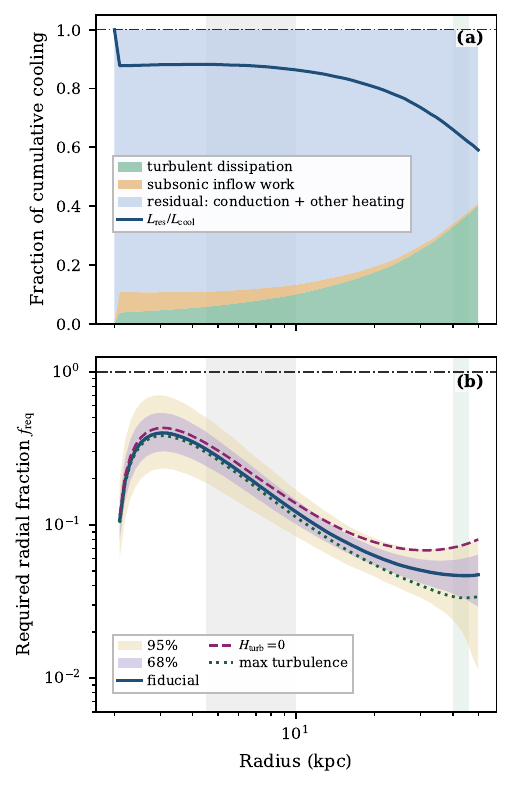}
\caption{\textit{Upper panel:} The median cumulative energy fractions is shown, where the green area is turbulent dissipation, the orange area is the subsonic inflow contribution, and the blue area is the residual power available to conduction and to unmodeled heating channels such as AGN work, mixing, waves, or shocks. The solid curve traces $L_{\rm res}/L_{\rm cool}$, which declines from $\simeq0.88$ in the inner core to $\simeq0.6$ at 50 kpc as turbulence becomes more important outward. \textit{Lower panel:} The spherical radial Spitzer fraction $f_{\rm req}$ that would be required in the maximal case where conduction carries the full residual. The gold and lavender bands show the 95\% and 68\% model ranges, while the dashed and dotted curves give the no-turbulence and maximum-turbulence brackets. The horizontal dot-dashed line marks full Spitzer conduction, $f_{\rm req}=1$. Grey and green vertical bands mark the inner and outer cold-front zones in both panels.}
\label{fig:energy}
\end{figure}

The explicit non-conductive terms do not close the Ophiuchus cooling budget. The subsonic inflow is negligible, with a median maximum radial Mach number of only $\simeq0.008$ and $L_{\rm comp}/L_{\rm cool}\simeq0.016$ at 25 kpc. The mean flow therefore redistributes thermal energy without materially altering the required heating. Turbulent dissipation is larger and becomes progressively more important outward. Even when the XRISM-anchored normalization is driven to its upper limit, the cumulative turbulent fraction is $\eta_{\rm turb}\simeq0.21$ at 25 kpc and reaches $\simeq0.34$ only in the most permissive accepted models. Figure~\ref{fig:energy}(a) shows the resulting separation directly. Roughly 60--90\% of the cumulative cooling remains after subtracting the modeled turbulent and inflow terms across 2--50 kpc. This is not assigned uniquely to conduction. It is the power available to conduction together with AGN work, mixing, waves, shocks, cosmic rays, and any other contribution represented by $L_{\rm other}$.

If conduction alone carries that residual, the radial coefficients required by the measured temperature gradient are
\begin{equation}
\begin{aligned}
 \freq(25\,\mathrm{kpc})&=\Ftwentyfive,\\
 \freq^{\rm in}&=\Finner,\qquad
 \freq^{\rm out}=\Fouter.
\end{aligned}
\label{eq:fresults}
\end{equation}
These values lie far below the full Spitzer coefficient. At 25 kpc, removing turbulent heating altogether raises the requirement only to $\freq=0.070$, while saturating the adopted turbulent upper limit lowers it to $0.046$. The inner-front requirement similarly spans only $0.166$--$0.197$ across the same bracket. The least certain explicit heating term therefore changes the normalization but does not control the conclusion. The decline of $\freq$ toward the outer core reflects a residual demand that falls faster than the conductive capacity associated with the still-positive temperature gradient. The radial solution is also safely classical, with $F_{\rm res}/q_{\rm sat}\simeq2\times10^{-3}$ near 25 kpc. The energy budget alone therefore permits a substantial conductive contribution at sub-Spitzer strength.

\subsection{\textcolor{black}{Cold-front survival limits the same transport}}\label{sec:res_fronts}

\begin{figure}[!t]
\centering
\includegraphics[width=0.84\columnwidth]{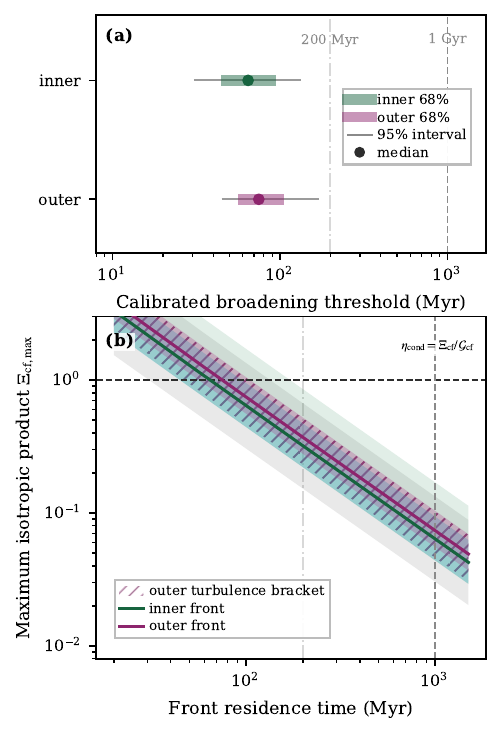}
\caption{\textit{Upper panel:} The numerically calibrated
broadening thresholds for the inner and outer fronts are shown, where the thick intervals show the 68\% ranges, the thin grey intervals show the 95\% ranges, and the points mark the medians. The vertical lines indicate 200 Myr and 1 Gyr, which serve as reference lifetimes rather than measured front ages. \textit{Lower panel:} The direct upper bound on $\Xicf=\Gcf\eta_{\rm cond}$ as a function of assumed residence time is shown. The hatched band spans the range permitted by the turbulent-heating normalisation at the outer front, while the colored percentile envelopes include the thermodynamic-profile uncertainty.}
\label{fig:fronts}
\end{figure}

The decisive test is whether those energetically viable coefficients can also operate normal to the two cold fronts. Applying the values of Equation~(\ref{eq:fresults}) to the calibrated diffusion problem gives passive broadening times to the published 99\% width limits of
\begin{equation}
 t_{\rm br}^{\rm in}=\Tinner\ \mathrm{Myr},\qquad
 t_{\rm br}^{\rm out}=\Touter\ \mathrm{Myr}.
\label{eq:tbr}
\end{equation}
These are broadening thresholds, not measurements of the front ages. They specify how long an initially sharp interface can remain narrower than the observed limit if the full-residual radial coefficient acts isotropically across it. The agreement between the two fronts is notable because they differ by almost an order of magnitude in radius and by more than a factor of three in their permitted widths (Figure~\ref{fig:fronts}a). The common result therefore points to a transport constraint rather than a peculiarity of either individual edge.

The thresholds are short compared with characteristic sloshing evolution times in MHD calculations \citep{ZuHone2011,ZuHone2013}. We do not interpret this comparison as a measurement of the residence time, since sloshing can regenerate and resharpen interfaces. Instead, it identifies the reference assumptions that cannot all hold simultaneously over a long exposure. Either conduction supplies less than the full residual, the front-containing sector carries less radial heat flux than the spherical mean, the interfaces are repeatedly sharpened, or the transport is intrinsically anisotropic. The front calculation constrains the combined effect of these possibilities rather than selecting one of them.

This distinction is important because a moving front need not broaden as a freely diffusing planar interface. In the idealized stagnation-flow problem of \citet{Xiang2007}, advection can establish a quasi-steady conductive layer with characteristic width $\Delta\sim(DR/U)^{1/2}$, where $D$ is the thermal diffusivity, $R$ the front curvature scale, and $U$ the incident flow speed. Our $t_{\rm br}$ therefore has a deliberately narrower meaning: it is the passive-diffusion exposure limit for the measured Ophiuchus interfaces. Sustained advective compression or repeated sloshing-driven sharpening would relax that limit and belongs explicitly to the dynamical-resharpening branch of the constraint above.

\subsection{The combined constraint on isotropic transport}\label{sec:res_combined}

The radial coefficient required by the energy budget, $\freq$, and the largest conductivity permitted normal to a front, $f_\perp^{\max}$, are evaluated on the same accepted atmospheres. Their ratio therefore removes much of the thermodynamic normalization and directly measures the tension between energetically useful radial transport and front survival. At the reference residence time $t_{\rm res}=200$ Myr,
\begin{equation}
 \Xi_{\rm cf,in}^{\max}=\Xiinner,\qquad
 \Xi_{\rm cf,out}^{\max}=\Xiouter,
\label{eq:xi200}
\end{equation}
so under the representative-sector hypothesis $\Gcf=1$, no more than about one third of the residual-carrying radial flux can pass normally through either interface. The corresponding transport contrast is
\begin{equation}
 \mathcal A_{\rm in}=3.1^{+1.4}_{-1.0},\qquad
 \mathcal A_{\rm out}=2.7^{+0.9}_{-0.8}.
\label{eq:aniso}
\end{equation}
\textcolor{black}{Here $\mathcal A\equiv\freq/f_\perp^{\max}$ is simply the ratio of the radial conductivity required to carry the residual cooling power to the largest conductivity permitted normal to the front. Thus $\mathcal A>1$ means that a single locally isotropic conductivity cannot satisfy both requirements.} The condition $\mathcal A>1$ is satisfied by 99.9\% of accepted models at the inner front and 98.8\% at the outer front. The factor-of-$\sim3$ mismatch is therefore not produced by the median atmosphere or by a narrow subset of the parameter space.

The 200 Myr value is a reference exposure, not a measured age. On this timescale, however, a sloshing cold front is not static. Continuous uplift of low-entropy gas toward the contact discontinuity and tangential advection along the front can replenish the thermodynamic contrast while conduction acts, and line-of-sight projection may further reduce the apparent signature of intrinsic broadening. The 200~Myr constraint should therefore be interpreted as a passive-broadening reference case rather than as a dynamical reconstruction of a front evolving for 200~Myr. Such dynamical maintenance or projection effects would relax the inferred limit on cross-front conductivity. Figure~\ref{fig:fronts}(b) shows the full residence-time dependence, with $\Xicf^{\max}$ decreasing monotonically as the assumed exposure grows. The turbulent normalization produces the largest systematic response. Under maximum turbulence the outer-front limit spans \XiouterSystematicRange\ and its transport contrast decreases to $\mathcal A_{\rm out}\simeq1.9$, with 2.5\% of accepted models falling below unity. The inner front is more robust, retaining $\mathcal A_{\rm in}\gtrsim2.9$ even under maximized turbulent heating. The result therefore separates into two broad physical regimes. If conduction supplies only a modest part of $L_{\rm res}$, other heating channels can dominate without violating the front widths. If conduction contributes substantially, the effective heat transport must be preferentially directed along the core temperature gradient rather than across the cold-front surfaces. What is excluded is the simultaneous combination of a large, long-lived, azimuthally representative, and locally isotropic conductive flux.

\section{Discussion and Conclusions}\label{sec:discussion}

The principal result is a direct comparison between two transport requirements that are usually evaluated separately. Energy-balance studies ask how much radial conductivity is needed to offset cooling, while cold-front studies ask how little cross-interface conductivity is required to preserve a sharp edge. Ophiuchus allows both quantities to be evaluated on the same gas. The radial temperature gradient can carry the residual power at sub-Spitzer strength, yet the two fronts permit only about one third of that full-residual transport to act normally across their surfaces at a representative 200 Myr exposure. The corresponding factor-of-$\sim3$ contrast is recovered independently at two radii and across nearly the entire accepted model ensemble. This is the key step beyond a conventional cold-front suppression estimate. The fronts constrain the conductivity that is energetically relevant to the core. \textcolor{black}{The energetic viability of conduction should not be interpreted as an alternative to feedback regulation. A conduction--cooling balance does not by itself provide the rapid self-regulation required to prevent thermal runaway, whereas condensation of low-entropy gas can couple changes in the cooling state to black-hole accretion and subsequent AGN heating \citep{Sharma2012,Prasad2015}. Our result therefore constrains the conductive contribution within a feedback-regulated cool core rather than proposing conduction as a replacement for AGN feedback.}

Magnetised heat transport offers a natural physical route to this directional behavior. Electrons conduct primarily along magnetic field lines in the weakly collisional intracluster medium \citep{Braginskii1965}, while sloshing shear can stretch and amplify the field into layers that approximately follow cold-front surfaces \citep{Lyutikov2006,ZuHone2011,ZuHone2013,ZuHone2021}. Such geometry can suppress the normal heat flux without requiring the entire core to be thermally disconnected. Anisotropic conduction is nevertheless not equivalent to perfect insulation. Sloshing MHD simulations show that field lines that are imperfectly draped can connect cold gas beneath a front to hotter regions elsewhere, so full Spitzer conduction along the field can substantially weaken the characteristic temperature jumps even when direct cross-front transport is suppressed \citep{ZuHone2013}. The very sharp Virgo front provides an observational counterpart: comparison with tailored simulations indicates that maintaining the measured discontinuities requires suppression of field-aligned conduction as well \citep{WernerVirgo2016}. The Ophiuchus radio data are qualitatively compatible with a structured magnetic environment but do not measure the field topology at either X-ray front. \citet{Murgia2010} inferred a model-dependent volume-averaged field of $0.3\,\mu$G under a specific inverse-Compton interpretation of the hard X-ray component, and MeerKAT later resolved narrow synchrotron threads that may trace magnetic structure, cosmic-ray density enhancements, or localized reacceleration \citep{Botteon2025}. Magnetic draping is therefore a plausible realization of the required anisotropy, not a unique explanation established by the present data. Three-dimensional field-line connectivity, kinetic suppression of parallel conduction, intermittency, azimuthal diversion of heat, and repeated dynamical sharpening can also reduce the effective normal flux.

\textcolor{black}{The dominant remaining uncertainty is geometric. In physical terms, the fronts constrain the product of the conductive share of the residual heating and the radial heat flux carried by the front-containing sector relative to the spherical mean; we denote this combination by $\Xicf=\Gcf\eta_{\rm cond}$.} If a front-containing sector carries less radial heat than the spherical mean, the locally isotropic share allowed by that front increases. Because $\Gcf$ enters multiplicatively and is not measured here, we report $\Xicf$ as the primary observable combination. The sector-averaging test addresses a different question. It shows that replacing the measured sectors with a spherical-equivalent atmosphere changes the conductive capacity at the $\lesssim25\%$ level, well below the factor-of-$\sim3$ transport contrast that motivates the result, while leaving the cooling emissivity essentially unchanged.

The remaining modeling assumptions are conservative in identifiable directions. The nonlinear front calculation evolves an intrinsic one-dimensional interface without curvature, line-of-sight projection, or the \textit{Chandra} PSF. Projection and instrumental broadening would make the intrinsic fronts narrower than the observed limits and would therefore reduce the permitted $f_\perp$. Conversely, sloshing-driven uplift and tangential advection can continuously restore the intrinsic thermodynamic contrast, so the mapping between observed width and conductive exposure is not unique. The XRISM forward model uses the live-pixel geometry with a varied CIE emissivity surrogate rather than a full response-folded spectral calculation. This affects the projection consistency test but not the radial energy balance that sets $\freq$. The turbulent closure is also deliberately permissive. Treating the XRISM-derived heating fraction as an upper limit minimizes the residual power assigned to conduction, while \citet{Russell2026} find turbulent dissipation to be insufficient to balance cooling by roughly a factor of three. A smaller turbulent contribution would increase $L_{\rm res}$ and strengthen the required radial-to-normal transport contrast.

\textcolor{black}{The broader implication follows from the steep temperature rise in Ophiuchus. The gas reaches $\sim9$~keV at tens of kiloparsecs, where the Spitzer coefficient is large because $\ksp\propto T^{5/2}$, while the cooler central gas provides the radial temperature gradient that drives heat inward.} Even in this favorable case, the radial conductivity capable of contributing substantially to the cooling balance cannot remain large, long-lived, azimuthally representative, and locally isotropic across the observed fronts. If conduction is subdominant here, the case for isotropic conductive heating becomes still weaker in cooler cores. If it is dynamically important, the Ophiuchus fronts turn that heating channel into a direct probe of directional heat transport in the intracluster medium. Deeper XRISM velocity mapping can reduce the dominant turbulent uncertainty at the outer front, while radio polarimetry across the interfaces can test whether the magnetic field is preferentially aligned with the front surfaces. The central conclusion is therefore not that conduction is absent, but that any substantial conductive contribution to cool-core heating must respect the geometry revealed by the cold fronts.

\begin{acknowledgments}
S.M. and R.S. acknowledge support from the Department of Atomic Energy, Government of India, under project no. RTI4019. N.W. was supported by GACR grant 21-13491X. The authors also thank Prateek Sharma and Kartick Chandra Sarkar for their valuable comments. This work made use of \texttt{OverCite} \citep{Shariat2026}, an in-editor citation tool for \LaTeX. Generative-AI tools were used for language editing and code debugging; the authors verified the scientific
content, calculations, references, and final text. 
\end{acknowledgments}

\software{NumPy, SciPy, Matplotlib}

\bibliographystyle{aasjournal}
\bibliography{references}
\end{document}